\documentclass[a4paper]{jpconf}
\usepackage{graphicx}

\usepackage{citesort}

\begin{document}

\title{TeV-PeV cosmic-ray anisotropy and local interstellar turbulence}

\author{Gwenael~Giacinti and John~G.~Kirk}

\address{Max-Planck-Institut f\"ur Kernphysik, Postfach 103980, D-69029 Heidelberg, Germany}

\ead{Gwenael.Giacinti@mpi-hd.mpg.de}

\begin{abstract}
We calculate the shape of the large-scale anisotropy of TeV--PeV cosmic-rays (CR) in different models of the interstellar turbulence. In general, the large-scale CR anisotropy (CRA) is not a dipole, and its shape can be used as a new probe of the turbulence. The $400$\,TeV and $2\,$PeV data sets of IceTop can be fitted with Goldreich-Sridhar turbulence and a broad resonance function, but other possibilities are not excluded. We then present our first numerical calculations of the CRA down to 3\,TeV energies in 3D isotropic Kolmogorov turbulence. At these low energies, the large-scale CRA aligns well with the direction of local magnetic field lines around the observer. In this type of turbulence, the CR intensity is flat in a broad region perpendicular to field lines. Even though the CRA is quite gyrotropic, we show that the local configuration of the turbulence around the observer does result in the appearance of weak, \lq\lq non-gyrotropic\rq\rq\/ small-scale anisotropies, which contain information on the local turbulence level.
\end{abstract}

\section{Introduction}

The flux of TeV--PeV cosmic-rays (CR) at Earth is anisotropic at a level of $\sim 10^{-3}$. See e.g.~\cite{DiSciascio:2014jwa,2017PrPNP..94..184A} for reviews. The direction of the large-scale CR anisotropy (CRA) is broadly consistent with that of the local interstellar magnetic field~\cite{Schwadron2014}, and its shape cannot be described by a pure dipole~\cite{Aartsen:2012ma}. For instance, IceCube and IceTop data at $\geq 100$\,TeV energies hint at a flattening of the CR intensity in directions perpendicular to local magnetic field lines~\cite{Giacinti:2016tld}. In the present work, we first report on how the shape of the large-scale CRA depends on the properties of the local interstellar turbulence within about a CR mean free path from Earth. We then illustrate and test the main assumptions of our theory~\cite{Giacinti:2016tld} by extending the numerical simulations of Ref.~\cite{2012PhRvL.109g1101G} down to TeV energies: we calculate the CRA in different realizations of 3D isotropic Kolmogorov turbulence for a realistic ratio of the CR gyroradius to the turbulence coherence length.

\section{Large-scale CRA and local interstellar turbulence}

The gyroradius of TeV--PeV CR ($\sim 10^{-4} - 10^{-1}$\,pc) is substantially smaller than the typical coherence length of the interstellar turbulence $l_{\rm c}\sim 1 - 10$\,pc~\cite{Haverkorn:2008tb}, and CR are expected to diffuse preferentially along magnetic field lines. The fact that the observed CRA points in the direction of the local magnetic field~\cite{Schwadron2014} corroborates this expectation. Assuming that CR undergo pitch-angle diffusion in a 1D magnetic flux tube of length $d \leq l_{\rm c}$, which contains the Earth, and assuming that the problem is stationary, we have shown in Ref.~\cite{Giacinti:2016tld} that the CRA at Earth is proportional to:
\begin{equation}
g(\mu) = \frac{\int_{0}^{\mu} {\rm d}\mu' \, \left(1-\mu'^{2}\right)/D_{\mu'\mu'}}{\int_{0}^{1} {\rm d}\mu' \, \left(1-\mu'^{2}\right)/D_{\mu'\mu'}} \,,
\label{CRA}
\end{equation}
where $\mu = \cos \theta$, $\theta$ is the pitch-angle (angle between the direction of the ordered magnetic field and the CR momentum), and $D_{\mu\mu}$ is the pitch-angle diffusion coefficient, assumed to be homogeneous in the flux tube. Eq.~(\ref{CRA}) is valid as long as the CR mean free path is smaller than a few times $d$, see~\cite{Giacinti:2016tld} for more details. The amplitude of the CRA depends on the {\it a priori} unknown value of the CR flux in the magnetic flux tube, but its shape does not. Therefore, we work with $g(\mu)$ which corresponds to the CRA with its amplitude renormalized to 1. Eq.~(20) of Ref.~\cite{Giacinti:2016tld} expresses $D_{\mu\mu}$ as a function of a resonance function $R_{n}$, for which we try the two following functions. One with a narrow, \lq\lq N\rq\rq\/, and one with a broad, \lq\lq B\rq\rq\/, resonance:
\begin{equation}
  R_{n}^{\rm N} = \frac{\tau^{-1}}{(k_\parallel v_\parallel - \omega + n \Omega)^{2} + \tau^{-2}} \;, {\rm ~~~and~~~}
  R_{n}^{\rm B} = \frac{\sqrt{\pi}}{\left|k_{\parallel}\right| v_{\perp} \delta\mathcal{M}_{\rm A}^{1/2}} 
\exp \left( - \frac{(k_\parallel v_\parallel - \omega + n \Omega)^{2}}{k_{\parallel}^{2} v_{\perp}^{2} \delta\mathcal{M}_{\rm A}}\right) \;,
\label{RF}
\end{equation}
where $k_{\parallel}$ is the parallel component of the wavevector, $v_{\parallel} = c \mu$, $v_{\perp} = c \sqrt{1-\mu^{2}}$, $\omega$ is the angular frequency of the waves, $\Omega$ the CR gyrofrequency, and $n=0,\pm 1$ in our calculations. For $R_{n}^{\rm N}$, the broadening of the resonance is assumed to be dominated by the Lagrangian correlation time of the turbulence, $\tau$, and for $R_{n}^{\rm B}$, by fluctuations of the parallel magnetic field strength, which is encapsulated in the parameter $\delta\mathcal{M}_{\rm A}<1$. We test two models of turbulence: Fast magnetosonic mode turbulence with an isotropic power spectrum $I_{\rm F}(k)\propto k^{-3/2}$~\cite{ChoLazarian2002} and Goldreich-Sridhar turbulence whose Alfv\'en and pseudo-Alfv\'en modes have the power spectrum $I_{\rm A,S}({\bf k})\propto k_{\perp}^{-10/3}\exp(-k_{\|}l^{1/3}/k_{\perp}^{2/3})$~\cite{Cho:2001hf}, where $l$ denotes the outer scale of the turbulence.

\begin{figure*}
\begin{center}
\includegraphics[width=0.32\textwidth]{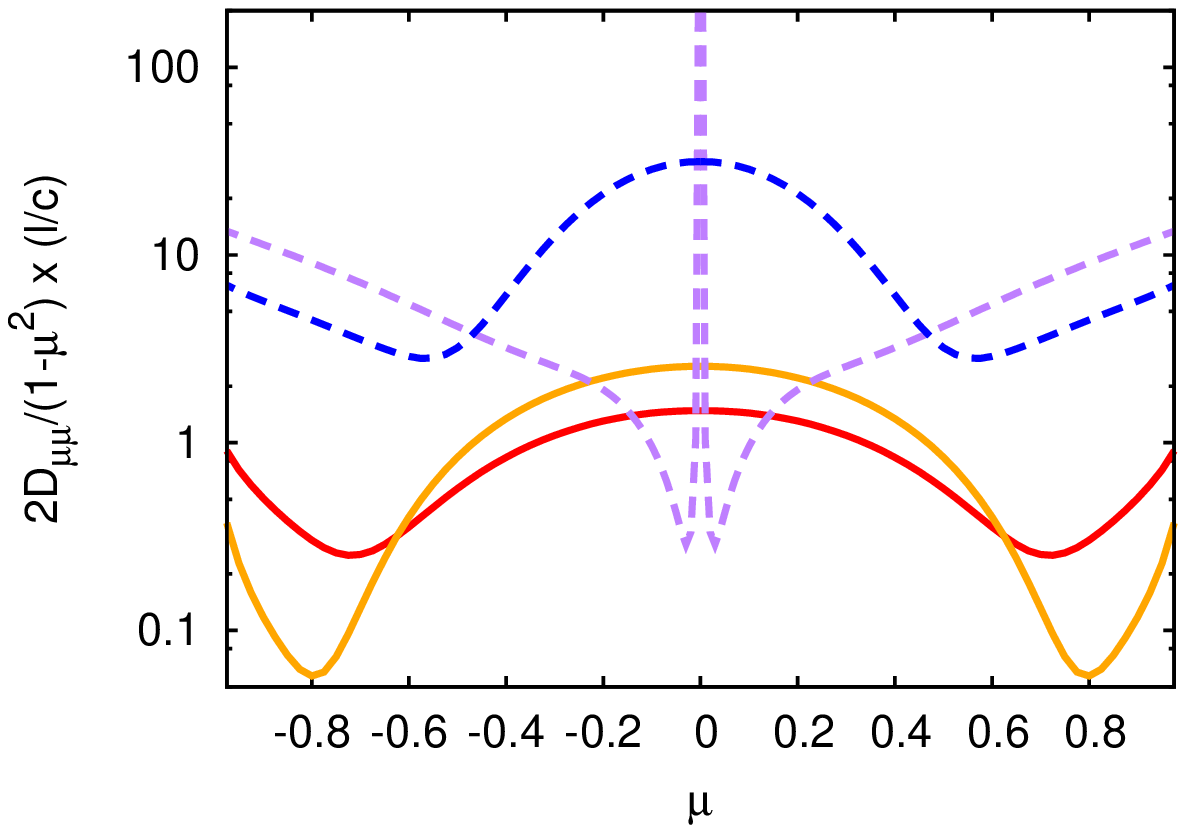}
\includegraphics[width=0.32\textwidth]{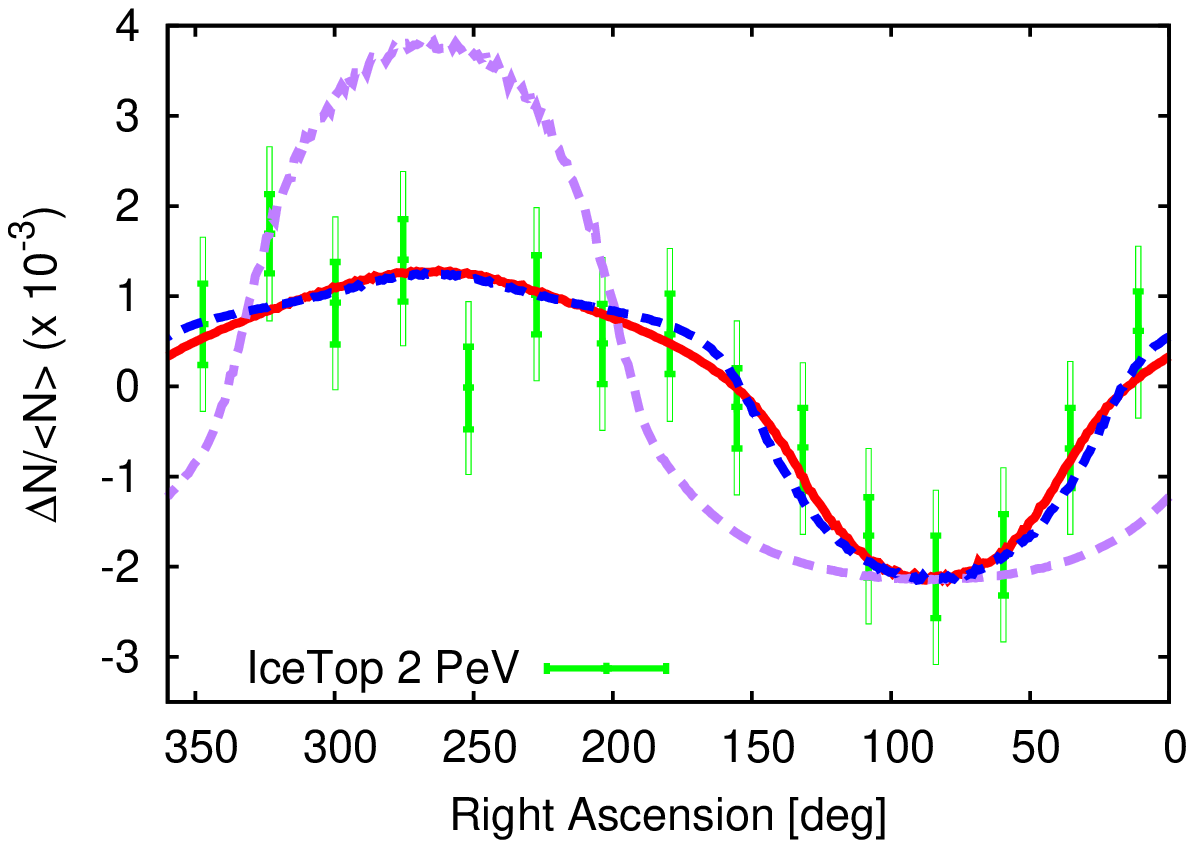}
\includegraphics[width=0.32\textwidth]{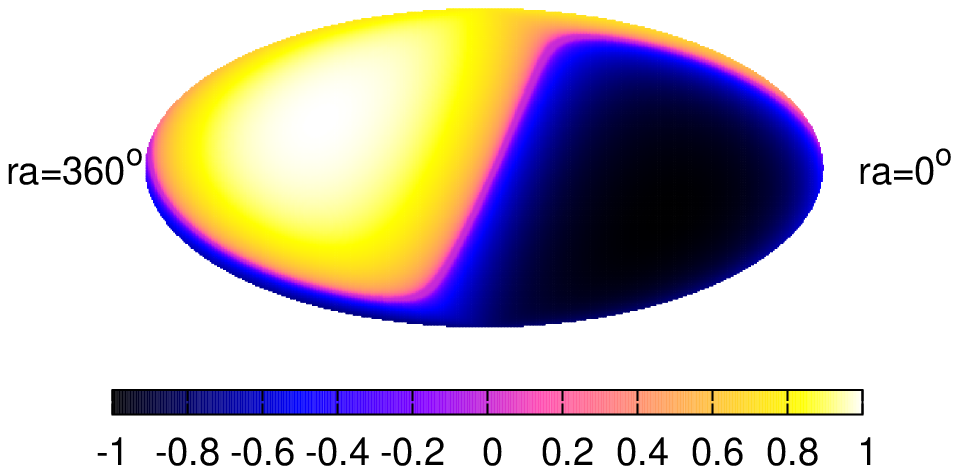}
\includegraphics[width=0.32\textwidth]{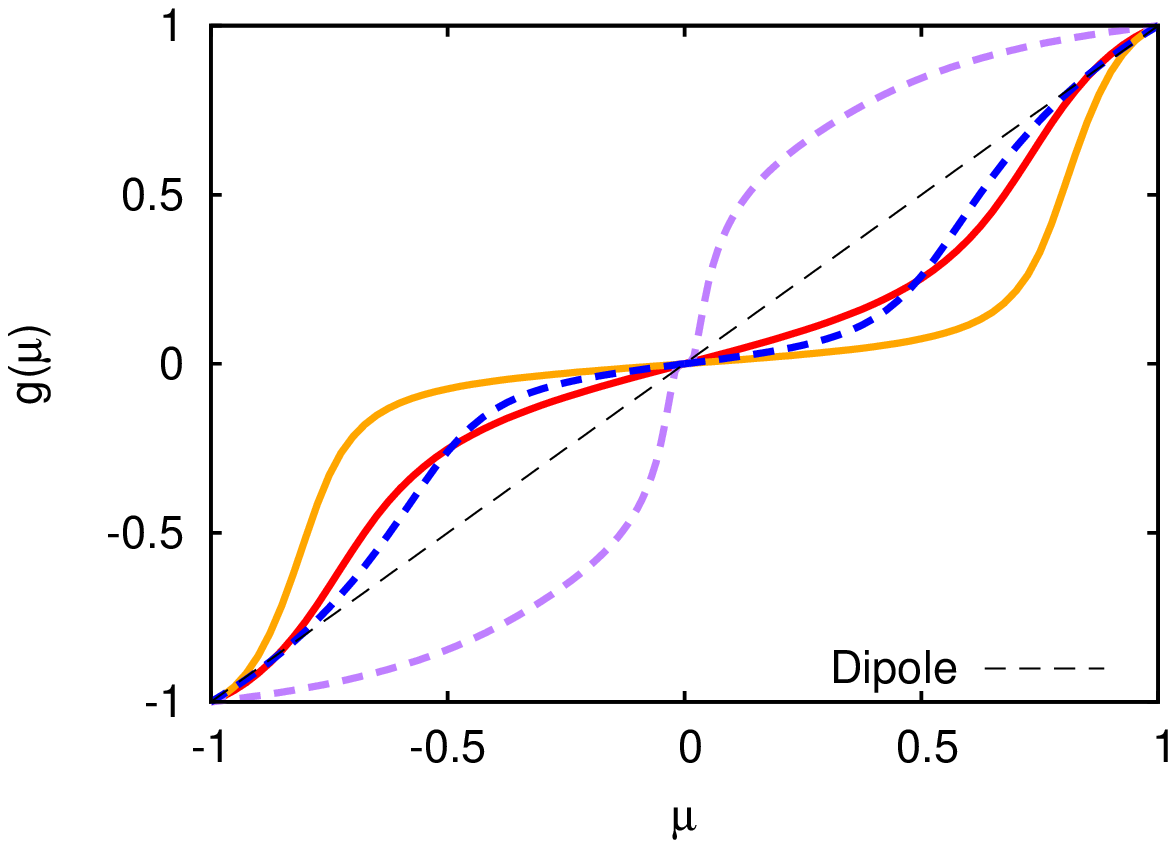}
\includegraphics[width=0.32\textwidth]{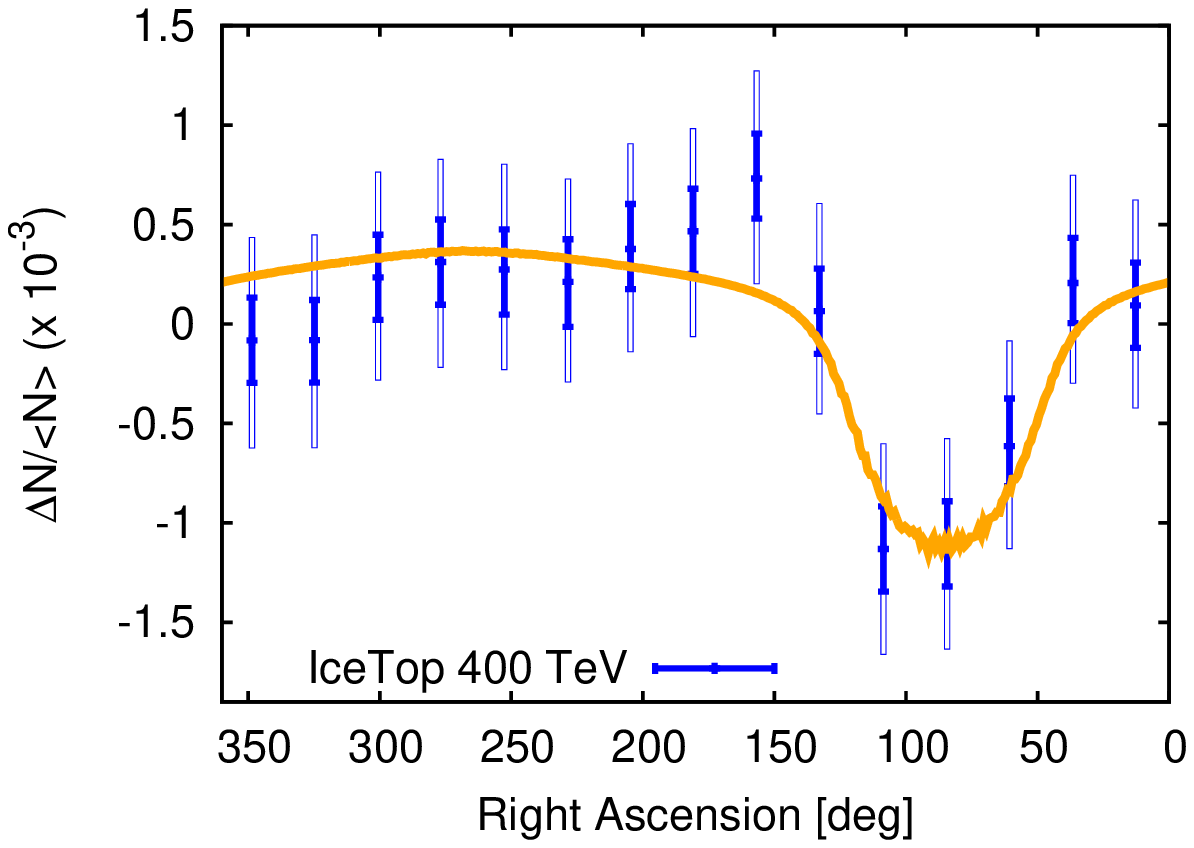}
\includegraphics[width=0.32\textwidth]{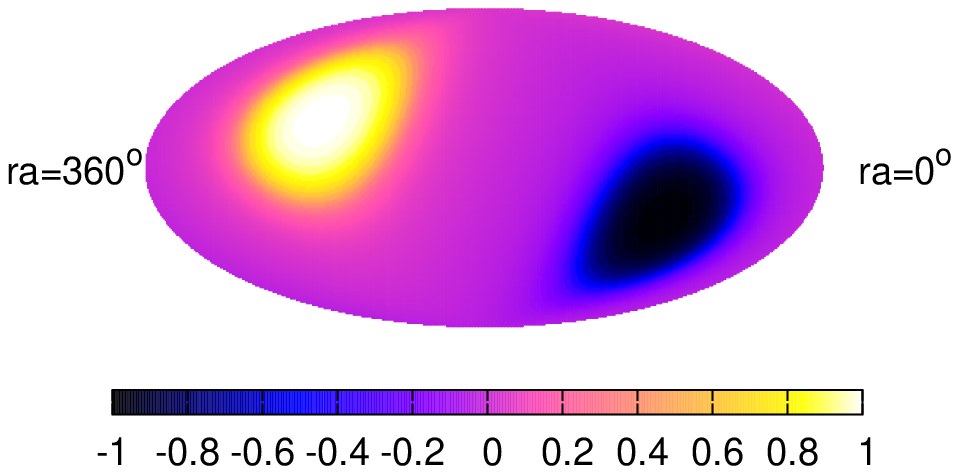}
\caption{\label{Fig1} Pitch-angle scattering rate $\nu(\mu)$ (upper left), anisotropy $g(\mu)$ (lower left), and relative CR intensity versus right ascension in the field of view of IceTop (middle column), for three models of the interstellar turbulence: Fast mode turbulence with $\epsilon = 10^{-3}$ and with either $R_{n}^{\rm N}$ and $u_{\rm A} = 10$\,km/s (dashed purple lines) or $R_{n}^{\rm B}$ and $\delta M_{\rm A}=0.1$ (dashed blue lines); Goldreich-Sridhar turbulence with $R_{n}^{\rm B}$, $\delta M_{\rm A}=0.33$, and $\epsilon = 10^{-2}$ (solid red lines) or $\epsilon = 10^{-3}$ (solid orange lines). Also shown are the 2\,PeV (upper middle) and 400\,TeV (lower middle) data taken from Ref.~\cite{Aartsen:2012ma}. Right column: CRA in equatorial coordinates for the dashed purple lines (upper panel), and the solid orange ones (lower panel).}
\end{center}
\end{figure*}

In Fig.~\ref{Fig1}, we present our calculations of the dimensionless pitch-angle scattering rate $\nu(\mu)=2D_{\mu\mu}/(1-\mu^2)\times(l/c)$ (upper left panel), and normalized large-scale CRA, $g(\mu)$ (lower left panel), for these models. The dashed lines are for fast mode turbulence, and the solid ones for Goldreich-Sridhar turbulence. In the middle column of Fig.~\ref{Fig1}, we plot the relative CR intensity versus right ascension in the field of view of IceTop experiment, which observes part of the Southern hemisphere, and compare with its 2\,PeV (top, green boxes) and 400\,TeV (bottom, blue boxes) data~\cite{Aartsen:2012ma}. The dashed purple lines correspond to fast mode turbulence with the narrow resonance function $R_{n}^{\rm N}$, and with $\tau$ calculated for an Alfv\'en velocity equal to $u_{\rm A} = 10$\,km/s ---see~\cite{Giacinti:2016tld} for more details. We set here the dimensionless CR rigidity to $\epsilon = c/(l\Omega) = 10^{-3}$. For this turbulence, $\epsilon$ only changes the normalization of $\nu(\mu)$, and not the shape of the CRA, $g(\mu)$. As can be seen in the upper left panel, $\nu$ has a very narrow peak around $\mu = 0$, which means that pitch-angle scattering is strongly enhanced for CR whose momenta are almost perpendicular to the local coherent field. This peak is due to the $n=0$ term in the expression for $D_{\mu\mu}$. At small $|\mu|$, and on both sides of the peak, $\nu$ goes through a minimum and recovers at $\mu \rightarrow \pm 1$. Since the derivative of $g(\mu)$ is proportional to $1/\nu$ (cf. Eq.~(\ref{CRA})), this results in a CRA that strongly varies at small values of $|\mu|$, and has broad, almost flat minima and maxima at $|\mu| \geq 0.5$, see the lower left panel. For comparison, the thin black dotted line shows $g(\mu)=\mu$ which corresponds to a dipole anisotropy. The anisotropy for the purple line is clearly not a dipole. For a better visualisation, we plot it in equatorial coordinates in the upper right panel of Fig.~\ref{Fig1}, assuming that its direction is given by that of the local coherent field measured by~\cite{Frisch:2012zj,Frisch:2015hfa}. By comparing with Fig.~6 of Ref.~\cite{Aartsen:2012ma}, one can see by eye that the shape of the CRA in the Southern hemisphere is very different from that measured by IceTop. This can also be seen in the upper middle panel, where we compare the resulting CR intensity with IceTop 2\,PeV data. In general, we find that models with $R_{n}^{\rm N}$ provide a bad match to the observations, which apparently require a flattening of $g$ in a moderately broad region around $\mu=0$, and hence a broad \lq\lq bump\rq\rq\/ in $\nu$ around $\mu=0$. $R_{n}^{\rm B}$ gives a better fit to the data: we show with dashed blue lines the results for fast modes with $R_{n}^{\rm B}$ and $\delta M_{\rm A}=0.1$. $\nu$ has a broad bump at $|\mu| < 0.5$ and minima at $\mu \simeq \pm 0.5$. This results in a CRA with a broad flat region at $|\mu| < 0.5$ and with two extrema whose angular half-widths are smaller than those of a dipole, see the lower left panel. This model provides a good fit to the 2\,PeV IceTop data, with a flat CR intensity at ${\rm RA}\geq 160^{\circ}$, and a small cold spot around ${\rm RA} \simeq 80 - 90^{\circ}$. However, the smaller angular size of the cold spot in the 400\,TeV data set cannot be well fitted with this model. If this change in the size of the cold spot with CR energy is real, this may point at an anisotropy in the power spectrum of the turbulence: If the anisotropy in Fourier space varies with $|{\bf k}|$, then CR with different energies interact with modes whose level of anisotropy is different. Goldreich-Sridhar turbulence is anisotropic, and the level of anisotropy of its modes in ${\bf k}$-space varies with $|{\bf k}|$. Solid lines show results for this turbulence with $R_{n}^{\rm B}$ and $\delta M_{\rm A}=0.33$. The red lines are for $\epsilon = 10^{-2}$, and the orange ones are for $\epsilon = 10^{-3}$. The ratio of these two values of $\epsilon$ is similar to the ratio of the energies of the two IceTop data sets --- the factor 2 difference does not affect our conclusions and is only due to the binninig in energy we chose for our scan of the parameter space. The scattering rate (upper left) is lower than for fast mode turbulence, but also presents broad maxima around $\mu=0$. The maximum is broader for $\epsilon = 10^{-3}$ than for $\epsilon = 10^{-2}$, and the CRA (lower left) flattens at $|\mu| < 0.7$ for $\epsilon = 10^{-3}$ and at $|\mu| < 0.5$ for $\epsilon = 10^{-2}$. The lower right panel shows the CRA for $\epsilon = 10^{-3}$ in equatorial coordinates. A small hot spot and a small cold spot are visible in the direction of the coherent magnetic field ($\mu=\pm 1$). They are separated by a wide magenta region, where the CR intensity is very flat. The two data sets from IceTop can be well fitted with this model of turbulence: The calculations for $\epsilon = 10^{-2}$ are compatible with the shape of the measured CR intensity at 2\,PeV (see the red line in the upper middle panel), and $\epsilon = 10^{-3}$ fits well the 400\,TeV data (see the orange line in the lower middle panel).

\section{Numerical simulations}

In the above analytical calculations, we made the assumption of gyrotropy. In this section, we investigate numerically the validity of this assumption. We propagate individual CR in different realizations of 3D isotropic Kolmogorov turbulence and calculate the CRA, using the backtracking method described in Ref.~\cite{2012PhRvL.109g1101G}. We consider here, for the first time, CR with energies as low as 3\,TeV. This is more than 3 orders of magnitude smaller than the energies probed in Ref.~\cite{2012PhRvL.109g1101G}, and, therefore, requires substantially longer calculation times. This value of 3\,TeV is in the relevant range for a direct comparison with the low energy measurements of the CRA at TeV energies. We note that, at even lower CR energies, time variations of the local turbulence would start to make small-scale features of the CRA vary on time scales smaller than the typical lifetime of an experiment, assuming local fluid and Alf\'en velocities of a few tens of km/s. We do not investigate this regime here. We use turbulence with root-mean-square strength $B_{\rm rms}=4\,\mu$G, and outer scale $l=150$\,pc, which corresponds to $l_{\rm c}=30$\,pc. The only difference between our numerical technique here and that of Ref.~\cite{2012PhRvL.109g1101G} is that we propagate here CR on finite distances $ct$ (as in Ref.~\cite{2015ApJ...815L...2A}), instead of stopping the trajectories on a sphere with a fixed radius. This difference is unimportant for the purpose of the present study. We propagate CR on distances of a few hundreds of pc, until the CRA converges. The angular shape of the CRA converges quickly, after about a CR mean free path. However, its absolute amplitude converges only after the CR have probed distances greater than a few coherence lengths of the turbulence. For shorter distances, the centre of mass of a set of initially nearby trajectories still strongly depends on the bending of local magnetic field lines around the observer.

\begin{figure*}
\begin{center}
\includegraphics[width=0.49\textwidth]{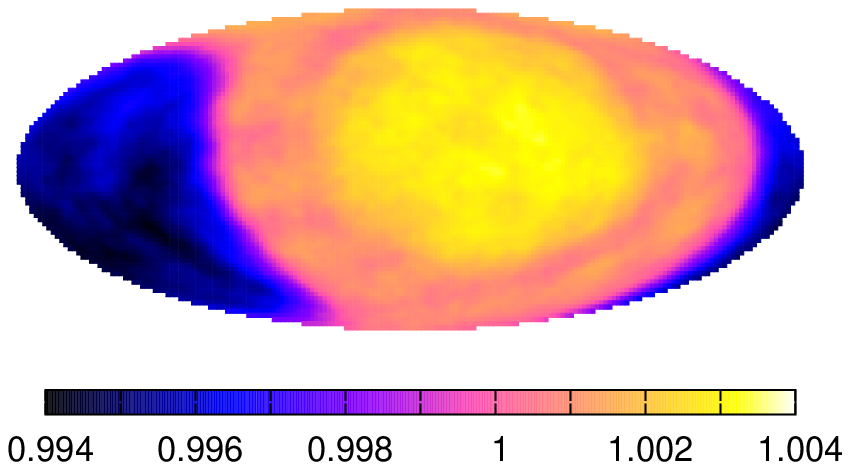}
\includegraphics[width=0.49\textwidth]{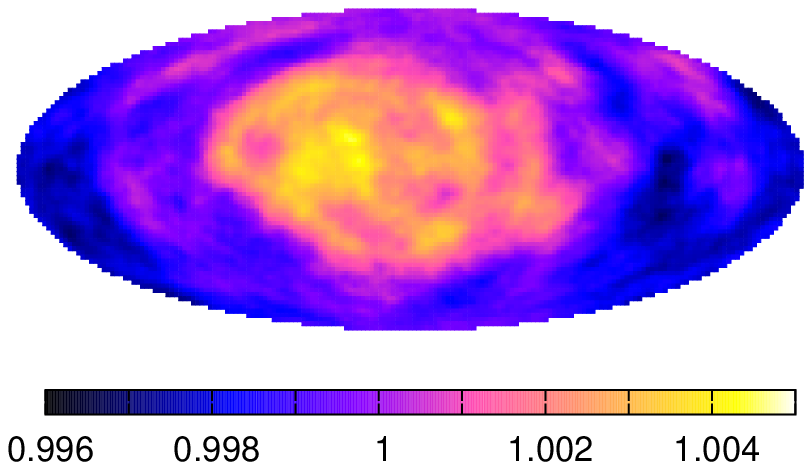}
\caption{\label{Fig2} Numerical calculations of the relative CR flux at 3\,TeV in two different realizations of 3D isotropic Kolmogorov turbulence with $B_{\rm rms}=4\,\mu$G and $l=150$\,pc. The CR intensity in these maps is averaged over $10^{\circ}$-radius circles.}
\end{center}
\end{figure*}

In Fig.~\ref{Fig2}, we show two sky maps of the relative CR flux (normalized to the angle-averaged flux) at 3\,TeV in two different realizations of the turbulence, and averaged over $10^{\circ}$-radius circles. As expected, the direction of the large-scale anisotropy is found to point in the direction of local magnetic field lines around the observer in both simulations. We do not add any regular field to the turbulence: the local coherent field is provided by the modes whose wavelengths are much larger than the CR gyroradius. The two panels in Fig.~\ref{Fig2} show two limiting cases. In the left panel, the local turbulence level on the scale of a 3\,TeV CR gyroradius, and within a CR mean free path from the observer, is quite small in this realization. The anisotropy looks very smooth and almost perfectly gyrotropic. In the right panel, the local turbulence level happens to be higher in that realization. The anisotropy looks less regular, with non-gyrotropic small-scale fluctuations that are well above the numerical noise level. This confirms that the mechanism suggested in~\cite{2012PhRvL.109g1101G} for creating small-scale anisotropies also works at energies of a few TeV. In all these simulations, the amplitude of \lq\lq non-gyrotropic\rq\rq\/ small-scale anisotropies is substantially smaller than that of the large-scale anisotropy. This is in line with observations, which suggest that the amplitude of small-scale anisotropies is about an order of magnitude smaller than that of the large-scale anisotropy. This also justifies the assumption of gyrotropy in the calculations of $g(\mu)$ in the previous section.

\begin{figure}[h]
\includegraphics[width=18pc]{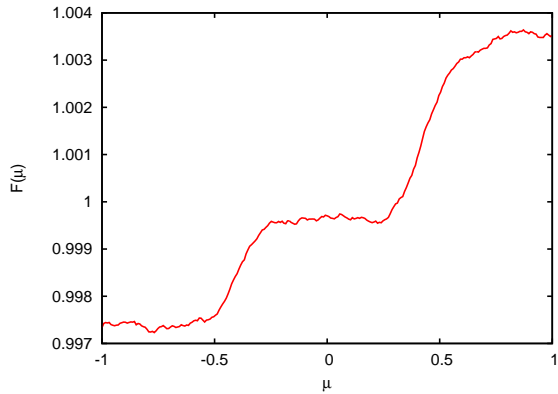}\hspace{2pc}%
\begin{minipage}[b]{14pc}\caption{\label{Fig3} Numerical calculation of the gyrophase-averaged relative CR flux $F(\mu)$, at 3\,TeV, for an observer in a given realization of 3D isotropic Kolmogorov turbulence with  $B_{\rm rms}=4\,\mu$G and $l=150$\,pc.}
\end{minipage}
\end{figure}

In Fig.~\ref{Fig3}, we plot our calculation of the gyrophase-averaged relative CR flux, $F$, versus $\mu$ for one of our simulations. The shape of $g(\mu)$ ($\propto F(\mu)-1$) shows some variations from one realization of the turbulence to another, but all cases we tested show a flattening of $g(\mu)$ at $|\mu|<0.25 - 0.5$. This tends to suggest that the CR pitch-angle scattering rate in 3D isotropic Kolmogorov turbulence may too have a broad bump around $\mu=0$.

\section{Discussion}

Both IceCube~\cite{Aartsen:2016ivj} and HAWC~\cite{2018ApJ...865...57A} have provided measurements of the angular power spectrum of the CRA. We have shown in Section~2 that the large-scale CRA $g(\mu)$ is not a pure dipole, except in the unphysical case of isotropic CR pitch-angle scattering where $D_{\mu\mu} \propto 1-\mu^{2}$. Therefore, a fraction of the power in the $C_{\ell}$'s, including those with $\ell \geq 2$, should be due to multipoles pointing roughly in the direction of local magnetic field lines. Among the several small-scale anisotropies detected by HAWC, one of them does not seem to vary substantially with CR energy: A small hot spot, located roughly in the direction of ${\rm RA} \sim 60^{\circ}$ and slightly below ${\rm dec} = 0^{\circ}$, remains at almost the same place on the sky from 2.0\,TeV to 72.8\,TeV, see Figs.~10 and 9 in Ref.~\cite{2018ApJ...865...57A}. When comparing with the Figs.~4 and 5 of Ref.~\cite{2018ApJ...865...57A}, it seems that the direction of this hot spot is quite roughly compatible with that of the large-scale anisotropy. We suggest here that this hot spot may be an additional sign that the \lq\lq large-scale\rq\rq\/ anisotropy pointing in the direction of magnetic field lines deviates from a dipole and contains high-order multipoles. This would naturally account for the apparent stability of this spot with energy. From a conceptual point of view, this suggestion and our study in Section~2 contain similarities with the suggestion of Ref.~\cite{Malkov:2010yq} for the origin of one of Milagro's hotspots, \lq\lq Region~A\rq\rq\/. However, our study differs in particular by the fact that we have used here the full general solution of the CR transport equation. We also note that the maximum significance of Milagro's Region~A is around ${\rm dec} \simeq 10^{\circ} - 20^{\circ}$ (see Fig.~1 of Ref.~\cite{2008PhRvL.101v1101A}), which is $\simeq 20^{\circ}$ away from the centre of HAWC's hot spot, and that Region A does not coincide with the direction of the maximum of the large-scale anisotropy as is visible in Fig.~2 of Ref.~\cite{2008PhRvL.101v1101A}. Some of these apparent differences may just be due to the different exposures of these experiments and data analysis techniques. A deeper study of these points would nonetheless be necessary in order to reach a firm conclusion on the origin of this particular hot spot in HAWC data.

The other small-scale anisotropies detected by HAWC, as well as those detected by IceCube, are not aligned with the large-scale anisotropy and require another explanation. They imply that the CR distribution at Earth is not perfectly gyrotropic. The mechanism discussed in Section~3, and initially presented in Ref.~\cite{2012PhRvL.109g1101G}, provides a natural explanation for them. Below a few TeV, heliospheric magnetic fields may too generate small-scale anisotropies for similar reasons, see e.g. Ref.~\cite{Zhang:2014dsu}. In any case, the fact that \lq\lq non-gyrotropic\rq\rq\/ small-scale anisotropies have been detected at 20\,TeV energy by IceCube~\cite{Aartsen:2016ivj} suggests that the mechanism of Ref.~\cite{2012PhRvL.109g1101G} should be at work, at least above several TeV.

\section{Conclusions}

Assuming pitch-angle diffusion of TeV--PeV CR in our local interstellar medium, we show that their large-scale anisotropy at Earth, $g(\mu)$, is in general not a dipole, see Section~2. $g(\mu)$ contains information on the local interstellar turbulence and CR propagation properties. Moderately broad resonance functions are favoured, and the 2\,PeV data set of IceTop can be fitted with isotropic fast modes or Goldreich-Sridhar turbulence. Thanks to its $|{\bf k}|$-dependent power spectrum, the latter type of turbulence can also explain the change in shape of the CRA between the 400\,TeV and the 2\,PeV data sets of IceTop. In Section~3, we present our first numerical calculations of the CRA in 3D isotropic Kolmogorov turbulence down to 3\,TeV. We find that $g(\mu)$ has a flattening in directions around $\mu = 0$ in this type of turbulence too. At these low energies, the CRA aligns well with the direction of local magnetic field lines around the observer and is quite gyrotropic. Weak, \lq\lq non-gyrotropic\rq\rq\/ small-scale anisotropies do nonetheless appear due to the local configuration of the turbulence around the observer at the time of the observations, as initially suggested in Ref.~\cite{2012PhRvL.109g1101G}. Their amplitude, which is always much smaller than that of the large-scale anisotropy $g(\mu)$, is connected to the turbulence level on resonant scales in our local magnetic flux tube.

\ack
This research was supported by a Grant from the GIF, the German-Israeli Foundation for Scientific Research and Development. GG also acknowledges useful discussions at the team meeting \lq\lq The Physics of the Very Local Interstellar Medium and Its Interaction with the Heliosphere\rq\rq\/, led by J.~Giacalone and J.~R.~Jokipii, and supported by the International Space Science Institute (ISSI) in Bern, Switzerland.

\section*{References}
\bibliography{references1}

\end{document}